\newlength{\absize}
\documentclass[12pt]{article}
\usepackage{graphicx}
\usepackage{latexsym}
\usepackage{amssymb}
\usepackage{bbm}
\usepackage{setspace}
\usepackage{braket}
\usepackage{bm}
\input prepictex
\input pictex
\input postpictex
\newdimen\tdim
\tdim=\unitlength
\def\stpltsmbl{\setplotsymbol ({\small .})}

\newbox\sru
\setbox\sru=\hbox{\beginpicture
\setcoordinatesystem units <\tdim,\tdim>
\stpltsmbl
\setquadratic
\plot
  0.0   0.0
  4.8   1.5
  7.5   5.0
  7.3   8.5
  5.0  10.0
  2.7   8.5
  2.5   5.0
  5.2   1.5
 10.0   0.0
/
\endpicture}
\def\springru #1 #2 *#3 /{\multiput {\copy\sru}  at
#1 #2 *#3 10 0 /}

\renewcommand{\bar}{\overline}

\newcommand{\spur}[1]{\!\not\! #1 \,}

\newcommand{\cA}{\mathcal{A}}

\newcommand{\cL}{\mathcal{L}}

\newcommand{\cO}{\mathcal{O}}

\newcommand{\cB}{\mathcal{B}}
\newcommand{\cC}{\mathcal{C}}
\newcommand{\cD}{\mathcal{D}}

\newcommand{\pd}{\partial}

\renewcommand{\slash}[1]{#1\!\!\!/}

\newcommand{\be}{\begin{equation}}
\newcommand{\ee}{\end{equation}}
\newcommand{\bea}{\begin{eqnarray}}
\newcommand{\eea}{\end{eqnarray}}
\newcommand{\ba}{\begin{array}}
\newcommand{\ea}{\end{array}}

\newcommand{\comment}[1]{}

\begin{document}

\thispagestyle{empty}
\pagestyle{empty}
\newcommand{\starttext}{\newpage\normalsize
 \pagestyle{plain}
 \setlength{\baselineskip}{3ex}\par
 \setcounter{footnote}{0}
 \renewcommand{\thefootnote}{\arabic{footnote}}
 }
\newcommand{\preprint}[1]{\begin{flushright}
 \setlength{\baselineskip}{3ex}#1\end{flushright}}
\renewcommand{\title}[1]{\begin{center}\LARGE
 #1\end{center}\par}
\renewcommand{\author}[1]{\vspace{2ex}{\large\begin{center}
 \setlength{\baselineskip}{3ex}#1\par\end{center}}}
\renewcommand{\thanks}[1]{\footnote{#1}}
\renewcommand{\abstract}[1]{\vspace{2ex}\normalsize\begin{center}
 \centerline{\bf Abstract}\par\vspace{2ex}\parbox{\absize}{#1
 \setlength{\baselineskip}{2.5ex}\par}
 \end{center}}

\title{Mass Perturbation Theory in the\\2-Flavor Schwinger Model with
Opposite Masses\\with a Review of the Background}
\author{
 Howard~Georgi\thanks{\noindent \tt hgeorgi@fas.harvard.edu}\\
Center for the Fundamental Laws of Nature\\
Jefferson Physical Laboratory \\
Harvard University \\
Cambridge, MA 02138\\
}
\date{\today}
\abstract{
I discuss the 2-flavor Schwinger model with $\theta=0$ and small equal and
opposite fermion masses (or $\theta=\pi$ with equal masses).  The massless
model has an unparticle sector with unbroken conformal symmetry.  I argue
that this special mass term modifies the conformal sector without breaking
the conformal symmetry.  I show in
detail how mass-perturbation-theory works for correlators of flavor-diagonal fermion 
scalar bilinears. The result provides quantitative evidence that the theory has no
mass gap for small non-zero fermion masses.  The massive fermions are bound
into conformally invariant unparticle stuff.  I show how the long-distance
conformal symmetry is maintained when small fermion masses are turned on
and calculate the relevant scaling dimensions for small mass.  I calculate
the corrections to the 2- and 4-point functions of the fermion-bilinear scalars to
leading order in perturbation theory in the fermion mass and describe a
straightforward procedure to extend the calculation to all higher
scalar correlators.  I hope that this model as a useful and non-trivial
example of unparticle 
physics, a sector with unbroken conformal symmetry coupled to interacting massive
particles, in which we can analyze the particle physics in a consistent
approximation.
}

\starttext
\section{Introduction\label{sec-intro}}
The massless 2-flavor Schwinger
model is an unparticle theory\footnote{See for example,
\cite{Georgi:2009xq}} in 1+1 dimensions
with a free massive scalar and a conformal 
sector that survives at low energy. 
In a previous paper,  \cite{Georgi:2020jik}, I discussed the massive
2-flavor Schwinger model, resolving some puzzles posed many years ago by
Coleman.~\cite{Coleman:1976uz}  Part of the resolution was a conjecture
that in the model with equal and opposite fermion masses at $\theta=0$ (or equal
masses at $\theta=\pi$), small fermion masses do not break the conformal
symmetry of the long-distance sector of the model 
even though the massive scalar has
nontrivial interactions.\footnote{For simplicity of presentation in this
paper, we will keep 
$\theta=0$.}  Thus I argued that the massive fermions are bound into
conformally invariant unparticle stuff. 
In this paper, I describe some quantitative evidence for this wild-sounding 
conjecture by finding the
correlation functions of the 
flavor-diagonal fermion-bilinear scalar conformal operators. I
find that the mass term does not break the conformal symmetry, but modifies
it and I calculate the  
non-trivial scaling dimensions of the unparticle stuff 
in perturbation theory in the fermion mass
parameter.  I introduce tools that make these calculations easier and
discuss some of the calculations in detail.

While I focus on the long-distance conformal theory in this paper, my
primary interest is in the particle physics of the full model.  I hope that
it is an example of unparticle physics, a conformal sector interacting with
massive particles without breaking the conformal symmetry,
with a well-defined procedure for calculation of physical
quantities.. 
Though the physics is still very simple, it is non-trivial and we can calculate.
The resulting theory may be
an interesting laboratory for studying the particle physics of
interacting unparticle theories.

\section{The Schwinger Model\label{sec-schwinger}}

The 
Lagrangian of the $n$-flavor Schwinger model is
\be
\cL=
\left( \sum_{j=1}^{n}\bar\psi_j\,\biggl(i\spur\pd - e\slash A\biggr)\,\psi_j\right)
-\frac{1}{4}F^{\mu\nu} F_{\mu\nu}
-\,\sum_{j=1}^{n}\mu_j\bar\psi_j\psi_j
\label{2f}
\ee
I begin by discussing $\mu_j=0$ and consider the mass term in
section~\ref{sec-ccmass}.\footnote{Some of ideas in this paper are related to
the analysis of diagonal color models in 1+1~\cite{Georgi:2019tch}.
See also
\cite{Belvedere:1978fj, GamboaSaravi:1981zd, Gattringer:1993ec,
Delphenich:1997ex}. }
\footnote{My conventions
are:  
$
g^{00}=-g^{11}=1,\,
\epsilon^{01}=-\epsilon^{10}=-\epsilon_{01}=\epsilon_{10}=1
$. From the defining properties $\{\gamma^\mu,\gamma^\nu\} = 2g^{\mu\nu}$ 
and $\gamma^5 = -\frac{1}{2}\epsilon_{\mu\nu}\gamma^\mu\gamma^\nu$, it follows that
$\gamma^\mu\gamma^5=-\epsilon^{\mu\nu}\gamma_\nu$ and
$\gamma^\mu\gamma^\nu=g^{\mu\nu}+\epsilon^{\mu\nu}\gamma^5$, and we will
use the representation 
$
\gamma^0=
\pmatrix{
0&1\cr
1&0\cr
},\;
\gamma^1=
\pmatrix{
0&-1\cr
1&0\cr
},\;
\gamma^5=\gamma^0\gamma^1=
\pmatrix{
1&0\cr
0&-1\cr
}\,$. Then in the massless theory, 
the Dirac components $\psi_1$ and $\psi_2$ describe right-moving and
left-moving fermions, respectively.}
The massless model has a classical $U(n)\times U(n)$ chiral symmetry acting
on the right- and left-moving fermion fields,
\begin{equation}
\psi_{j1}\equiv \frac{1+\gamma^5}{2}\psi_j\to R_{jk}\,\psi_{k1}
\quad
\psi_{j2}\equiv \frac{1-\gamma^5}{2}\psi_j\to L_{jk}\,\psi_{k2}
\label{chiral}
\end{equation}
It is broken
by the anomaly down to
$SU(n)\times SU(n)\times U(1)$.

In Lorenz gauge, $\partial_\mu A^\mu=0$, we can write
\be
A^\mu = \epsilon^{\mu\nu}\pd_\nu\cA/m
\label{A-decomposition}
\ee
where
\begin{equation}
m^2=n\,e^2/\pi
\label{m2-1}
\end{equation}
At this point, it is not obvious why we should choose $m$ this way but we
will see that the answer is the chiral $U(1)$ anomaly.  
Then the Lagrangian is
\be
\displaystyle
\cL =\left(\sum_{j=1}^{n}\biggl( i\bar\psi_j\spur\pd\psi_j -
e\bar\psi_j\gamma_\mu\psi_j
\epsilon^{\mu\nu}\pd_\nu\cA/m\biggr)\right)
 + \frac{1}{2m^2}\cA\,\Box^2\cA
\label{Sommerfield-AV}
\ee

If we change the fermionic variables to
\be
\Psi_j = e^{ie\cA\gamma^5/m}\psi_j
=e^{i(\pi/n)^{1/2}\cA\gamma^5}\psi_j
\label{psi-redef}
\ee
the fermions becomes free and the Lagrangian becomes
\be
\cL=\left(\sum_{j=1}^{n}i\bar\Psi_j\spur\pd\Psi_j\right) + 
\frac{1}{2m^2}\cA\,\Box^2\cA - \frac{1}{2}\pd_\mu\cA\pd^\mu\cA
\label{Sommerfield-redefined}
\ee
The last term is the effect of the anomaly.
It is worth recalling how this works in more detail.
The redefinition (\ref{psi-redef}) is an axial $U(1)$ transformation
--- $\pd^\mu\cA$ has
axial-vector couplings because 
\begin{equation}
\gamma_\mu
\epsilon^{\mu\nu}\pd_\nu\cA
=\gamma_\mu\gamma^5\pd^\mu\cA
\end{equation}
and an axial transformation induces a change in the Lagrangian because of
the chiral $U(1)$ anomaly.   The
effect from an infinitesimal axial transformation is proportional to the 2D
anomaly of the axial $U(1)$ current,
\begin{equation}
\pd_\mu j_5^\mu=-n\,\frac{e}{\pi}\epsilon^{\mu\nu}\pd_\mu A_\nu
=-n\,\frac{e}{\pi}\Box\cA/m
\label{anomaly}
\end{equation}
\be
\frac{d}{d\alpha}
\left(\sum_{j=1}^{n} \overline{e^{ie\alpha\cA\gamma^5/m }\psi_j}
\,\gamma_\mu\biggl(i\pd^\mu
 -e\,
\epsilon^{\mu\nu}\pd_\nu(1-\alpha)\cA/m\biggr)\,
e^{ie\alpha\cA\gamma^5/m }\psi_j\right)
=-\frac{n\,e^2}{m^2\pi}\cA(1-\alpha)\Box\cA
\label{Sommerfield-AV-ke-vda}
\ee
Integrating (\ref{Sommerfield-AV-ke-vda}) from $\alpha=0$ to $1$ gives
\begin{equation}
\left(\sum_{j=1}^{n} \bar\psi_j\,\gamma_\mu\biggl(i\pd^\mu
 -e\,
\epsilon^{\mu\nu}\pd_\nu\cA/m\biggr)\,
\psi_j\right)
=\left(\sum_{j=1}^{n}i\bar\Psi_j\spur\pd\Psi_j\right)
+\frac{n\,e^2}{2m^2\pi}\cA\Box\cA
=i\bar\Psi\spur\pd\Psi
-\frac{1}{2}\pd_\mu\cA\pd^\mu\cA
\label{anomaly2}
\end{equation}
where $\Psi$
is given by (\ref{psi-redef}). This is why we chose $m$ the way we did in 
(\ref{A-decomposition}) and (\ref{m2-1}).

Focusing on $\cA$ in (\ref{Sommerfield-redefined}), we can replace it with
somewhat more normal looking fields 
as follows.
\be
\frac{1}{2m^2}\cA\,\Box^2\cA - \frac{1}{2}\pd_\mu\cA\pd^\mu\cA
\to
-\frac{m^2}{2}\cB^2+\cB\Box\cA - \frac{1}{2}\pd_\mu\cA\pd^\mu\cA
\ee
\be
=-\frac{m^2}{2}\cB^2+\frac{1}{2}\pd_\mu\cB\pd^\mu\cB
- \frac{1}{2}\pd_\mu\cC\pd^\mu\cC
\ee
where 
\be
\cC=\cA+\cB
\ee
so $\cB$ is a massive free field
and $\cC$ is a massless ghost and
the Lagrangian becomes
\be
\cL=\left(\sum_{j=1}^{n}i\bar\Psi_j\spur\pd\Psi_j\right)
-\frac{m^2}{2}\cB^2+\frac{1}{2}\pd_\mu\cB\pd^\mu\cB
- \frac{1}{2}\pd_\mu\cC\pd^\mu\cC\\
\label{Sommerfield-redefined2}
\ee

Thus for gauge invariant correlators of 
local fields, the result of summing the perturbation theory to all orders
can be found simply by making the following replacements:\footnote{This
argument appears in 
\cite{Georgi:2019pje}.} 
\be
A^\mu = \epsilon^{\mu\nu}\pd_\nu(\cB-\cC)/m
\label{f-A-decomposition}
\ee
\be
F^{01}=\partial_\mu\partial^\mu(\cB-\cC)/m
\ee
\be
\psi_j = e^{-i(\pi/n)^{1/2}\,(\cC-\cB)\gamma^5}\Psi_j
\label{f-psi-redef-vcb}
\ee
with $m=e\sqrt{n/\pi}$ from (\ref{m2-1})
and using the free-field Lagrangian, (\ref{Sommerfield-redefined2}).

We will be particularly concerned with flavor-diagonal fermion-bilinear scalar
operators.
\begin{equation}
\cO_j=\psi_{j1}^* \psi_{j2}
=e^{2i(\pi/n)^{1/2}\,(\cC-\cB)}\Psi_{j1}^* \Psi_{j2}
\label{oj}
\end{equation}

The free-fermion bilinears (\ref{oj}) have the remarkable property of
``bosonization.''~\cite{Coleman:1974bu,Mandelstam:1975hb}  
For us, what this means is that any
non-zero
correlator of the $\cO_j$s and $\cO_j^*$s 
can be calculated in terms of the massive scalar field, $\cB$, the ghost
$\cC$, and free canonically normalized massless
``scalar fields'', $\cD_j$ with the replacement
\begin{equation}
\cO_j\to\frac{\xi m}{2\pi}\,e^{2i(\pi/n)^{1/2}\,(\cC-\cB)}\,e^{2i\pi^{1/2}\,\cD_j}
\end{equation}
where
\be
\xi\equiv e^{\gamma_E}/2\mbox{~~where $\gamma_E$ is Euler's constant.}
\label{xi}
\ee
Note that in perturbation theory, the only non-zero correlators are those
equal number of $\cO_j$s and $\cO_j^*$s for each $j$. 
But there are important non-perturbative effects,
again related to the anomaly.

The non-perturbative effects are particularly simple in the
1-flavor model, where (\ref{f-psi-redef-vcb}) gives
\be
\psi = e^{-i\pi^{1/2}\,(\cC-\cB)\gamma^5}\Psi
\label{f-psi-redef-vcb1}
\ee
and there is only one conjugate pair of scalar fermion bilinears
\begin{equation}
\cO_1=\psi_{1}^* \psi_{2}
\to e^{2i\pi^{1/2}\,(\cC-\cB)}\Psi_{1}^* \Psi_{2}
\to\frac{\xi m}{2\pi}\,e^{2i\pi^{1/2}\,(\cC-\cB)}\,e^{2i\pi^{1/2}\,\cD}
\label{o1n=1}
\end{equation}
Now the effects of the bosonization field $\cD$ are exactly canceled by the
effects of the ghost $\cC$ and (\ref{o1n=1}) is
\begin{equation}
\cO_1=\psi_{1}^* \psi_{2}
\to\frac{\xi m}{2\pi}\,e^{-2i\pi^{1/2}\,\cB}
\label{o1n=1b}
\end{equation}
Because the $\cB$ field is massive, this means that the effects of the
$\cO$ operators on one another are exponentially suppressed at distances
larger that $1/m$.  But then if we have any combination of $\cO_1$ and
$\cO_1^*$ fields in some region of space, we can look at their correlator
with a conjugate set in a distant region.  We can then calculate the
correlator perturbatively using (\ref{o1n=1b}) and as the distance between
the regions goes to infinity, the result factors
into a product of correlators in the separate regions.  Cluster
decomposition then implies that we can calculate the correlator of any
combination of $\cO_1$s and $\cO_1^*$s using (\ref{o1n=1b}) up to a phase
factor
\begin{equation}
\cO_1
\to e^{i\theta}\,\frac{\xi m}{2\pi}\,e^{-2i\pi^{1/2}\,\cB}
\label{o1n=1theta}
\end{equation}
This implies, among other things, that
\be
\braket{0|\cO_1|0}=e^{i\theta}\,\frac{\xi m}{2\pi}
\ee
so $\cO_1$ has a VEV that breaks the
chiral symmetry.
One can think of the massless field $\cD$ as the Goldstone
boson of the spontaneously broken chiral symmetry, but it is unphysical
because its effects are completely canceled by the ghost field $\cC$.\footnote{See
\cite{Kogut:1974kt}.
}
If we add a fermion mass, the $\cB$ field is no longer free and in addition
to the physical fermion mass, the parameter $\theta$ in (\ref{o1n=1theta})
becomes the physical $\theta$-parameter.\footnote{See for
example, \cite{Coleman:1976uz}.}

\section{Two flavors\label{sec-2-flavors}}

The 2-flavor model has a non-Abelian chiral symmetry, but we will again be mostly
concerned with the physics of the flavor diagonal fermion-bilinear scalars that carry
the chiral $T_3$ symmetry
\begin{equation}
\psi_{11}\to e^{i\phi}\psi_{11}
\quad
\psi_{21}\to e^{-i\phi}\psi_{21}
\quad
\psi_{12}\to e^{-i\phi}\psi_{12}
\quad
\psi_{22}\to e^{i\phi}\psi_{22}
\label{chiral-t3}
\end{equation}
and the chiral $U(1)$ symmetry
\begin{equation}
\psi_{11}\to e^{i\phi}\psi_{11}
\quad
\psi_{21}\to e^{i\phi}\psi_{21}
\quad
\psi_{12}\to e^{-i\phi}\psi_{12}
\quad
\psi_{22}\to e^{-i\phi}\psi_{22}
\label{chiral-u1}
\end{equation}

Now with two flavors, we can again write the (flavor-diagonal) fermion
bilinears in bosonized form
\begin{equation}
\psi_{j1}^* \psi_{j2}
=e^{i\sqrt{2\pi}\,\,(\cC-\cB)}\Psi_{j2}^* \Psi_{j1}
=\frac{\xi m}{2\pi}\,e^{i\sqrt{2\pi}\,\,(\cC-\cB)}\,e^{2i\pi^{1/2}\,\cD_j}
\label{boson2}
\end{equation}
and when we calculate any correlator that is non-zero in perturbation
theory, and thus allowed by the perturbatively conserved chiral symmetries
(\ref{chiral-t3}) and (\ref{chiral-u1}), standard bosonization arguments imply
that (\ref{boson2}) gives the result of summing the perturbation theory to all
orders.  

The massless scalar fields $\cD_1$ and $\cD_2$ and the ghost field $\cC$ do
not make physical  
sense in isolation because
of infrared divergences,~\cite{Coleman:1973ci} but their exponentials in
(\ref{boson2}) generate 
conformally invariant correlators in the theory at long distances and
the scale is fixed by the mass $m$. This is unparticle stuff with no
particle interpretation.~\cite{Georgi:2009xq}
For example
\begin{equation}
\braket{0|T\,
O_1(x)\,
O_1^* (y)
|0}
=\frac{(\xi
m)}{(2\pi)^2}
\exp
\left[
K_0\left(m\sqrt{-(x-y)^2 + i\epsilon}\right)\right]
\,\left(-(x-y)^2+i\epsilon\right)^{-1/2}
\label{11*}
\end{equation}
where
$K_0$ is related to the scalar propagator\footnote{Note that there is no
arbitrariness here because 
these composite operators do not require multiplicative renormalization for
$\mu_j=0$ 
so the position-space correlators are well-defined for non-zero
separation.  A subtractive renormalization is required for the 2-point
function at zero separation and is needed to define the Fourier transforms.}
\begin{equation}
K_0\left(m\sqrt{-x^2+i\epsilon}\right)=2\pi i\int\frac{d^2p}{(2\pi)^2}\,
\frac{e^{-ipx}}{p^2-m^2+i\epsilon}
\label{k0}
\end{equation}
(\ref{11*}) implies that 
the fermion bilinears have scaling dimension $1/2$ rather than their
naive engineering dimension $1$.  For $n>2$, the fermion bilinears have
scaling dimension $1-1/n$.  This is 
zero for $n=1$ which is why there is no conformal sector at all in the
Schwinger model.

We can rewrite (\ref{boson2}) as
\begin{equation}
\begin{array}{l}
\displaystyle
\psi_{11}^* \psi_{12}
=\frac{\xi m}{2\pi}\,e^{i\sqrt{2\pi}\,\,(\cC-\cB)}\,e^{i\sqrt{2\pi}\,\,(\cD_{+}+\cD_{-})}
=\frac{\xi m}{2\pi}\,e^{i\sqrt{2\pi}\,\,(\cC+\cD_+-\cB+\cD_-)}
\\
\displaystyle
\psi_{21}^* \psi_{22}
=\frac{\xi m}{2\pi}\,e^{i\sqrt{2\pi}\,\,(\cC-\cB)}\,e^{i\sqrt{2\pi}\,\,(\cD_{+}-\cD_{-})}
=\frac{\xi m}{2\pi}\,e^{i\sqrt{2\pi}\,\,(\cC+\cD_+-\cB-\cD_-)}
\\
\displaystyle
\psi_{12}^* \psi_{11}
=\frac{\xi m}{2\pi}\,e^{-i\sqrt{2\pi}\,\,(\cC-\cB)}\,e^{-i\sqrt{2\pi}\,\,(\cD_{+}+\cD_{-})}
=\frac{\xi m}{2\pi}\,e^{-i\sqrt{2\pi}\,\,(\cC+\cD_+-\cB+\cD_-)}
\\
\displaystyle
\psi_{22}^* \psi_{21}
=\frac{\xi m}{2\pi}\,e^{-i\sqrt{2\pi}\,\,(\cC-\cB)}\,e^{-i\sqrt{2\pi}\,\,(\cD_{+}-\cD_{-})}
=\frac{\xi m}{2\pi}\,e^{-i\sqrt{2\pi}\,\,(\cC+\cD_+-\cB-\cD_-)}
\end{array}
\label{tempted0}
\end{equation}
where
\begin{equation}
\cD_{\pm}=\frac{1}{\sqrt{2}}\left(\cD_1\pm\cD_2\right)
\end{equation}
Now $\cD_+$ transforms like a Goldstone boson associated with
spontaneous breaking of the chiral $U(1)$ and as in the $n=1$ model
its effects in gauge
invariant matrix elements are completely canceled by the ghost field
$\cC$.  
Thus we are tempted to write

\begin{equation}
\begin{array}{ll}
\displaystyle
O_1\equiv \psi_{11}^* \psi_{12}
\to\frac{\xi m}{2\pi}\,e^{-i\sqrt{2\pi}\,\,(\cB-\cD_-)}
&
\displaystyle
O_2\equiv \psi_{21}^* \psi_{22}
\to\frac{\xi m}{2\pi}\,e^{-i\sqrt{2\pi}\,\,(\cB+\cD_-)}
\\
\displaystyle
O_1^* \equiv \psi_{12}^* \psi_{11}
\to\frac{\xi m}{2\pi}\,e^{-i\sqrt{2\pi}\,\,(-\cB+\cD_-)}
&
\displaystyle
O_2^* \equiv \psi_{22}^* \psi_{21}
\to\frac{\xi m}{2\pi}\,e^{-i\sqrt{2\pi}\,\,(-\cB-\cD_-)}
\end{array}
\label{tempted}
\end{equation}
But (\ref{boson2}), (\ref{tempted0}), and (\ref{tempted}) cannot be right
in general because they would imply VEVs for  
$O_1$ and $O_2$,  and their conjugates,
breaking the chiral $T_3$  
symmetry spontaneously.  This cannot happen in 1+1
dimensions.~\cite{Coleman:1973ci}  

But (\ref{tempted}) is nevertheless a very useful shorthand because
we can show using cluster decomposition that it
gives the correct matrix elements for the correlators
that are not forbidden by
the conserved chiral $T_3$ symmetry, (\ref{chiral-t3}), up to the
arbitrary angle $\theta$.  To understand this,
note that we know that (\ref{tempted}) works for perturbatively allowed
correlators and consider two similar looking correlators
\begin{equation}
\braket{0|T\,
O_1(x)\,
O_2 (0)\,
O_1^*(y)\,
O_2^*(y+z)
|0}
\label{cxyz}
\end{equation}
and
\begin{equation}
\braket{0|T\,
O_1(x)\,
O_2^*(0)\,
O_1^*(y)\,
O_2(y+z)
|0}
\label{cxyz*}
\end{equation}
\begin{equation}
\mbox{for $-(y)^2\to\infty$ with $x^2$ and $z^2$ fixed}
\label{xyz}
\end{equation}

Cluster decomposition requires that in the limit (\ref{xyz}),
(\ref{cxyz}) factorizes into
\begin{equation}
\braket{0|T\,
O_1(x)\,
O_2 (0)
|0}
\,\braket{0|T\,
O_1^*(y)\,
O_2^*(y+z)
|0}
\label{cxyzf}
\end{equation}
When we calculate (\ref{cxyz}) using (\ref{tempted}), the exponentials of $K_0$
in
the terms that involve $y$ all go to 1 and the power-law terms are
\begin{equation}
\frac{\left(-y^2\right)^{1/2}\left(-(y+z-x)^2\right)^{1/2}}
{\left(-(y-x)^2\right)^{1/2}\left(-(y+z)^2\right)^{1/2}}\to 1
\label{xyzfrac}
\end{equation}
and we must conclude that the two factors in (\ref{cxyzf}) are non-zero and
the non-zero result is given by their calculation from
(\ref{tempted}) up an arbitrary phase $e^{i\theta}$.

But for (\ref{cxyz*})  the power law terms are
\begin{equation}
\left(-y^2\right)^{-1/2}\left(-(y+z-x)^2\right)^{-1/2}
\left(-(y-x)^2\right)^{-1/2}\left(-(y+z)^2\right)^{-1/2}\to 0
\label{xyzfrac*}
\end{equation}
consistent with the fact that expectation values in the separate factors vanish
because of chiral $T_3$ conservation.

Similar considerations apply to all correlators and we can
use (\ref{tempted}) to calculate all correlators with zero chiral $T_3$
up to a single arbitrary phase that we will set to 1.  When we add a mass
term in the next section, this means that will keep $\theta=0$.

So for example
\begin{equation}
\braket{0|T\,
O_1(x)\,
O_2(y)
|0}
=\frac{(\xi
m)}{(2\pi)^2}
\exp
\left[
-K_0\left(m\sqrt{-(x-y)^2 + i\epsilon}\right)\right]
\,\left(-(x-y)^2+i\epsilon\right)^{-1/2}
\label{12}
\end{equation}

The dictionary for computing the non-zero correlators of the
exponentials is standard.  
Between each pair of operators we include the terms
\begin{equation}
\braket{0|e^{is_1\sqrt{2\pi} \,\cB(x)}\,e^{is_2\sqrt{2\pi} \,\cB(y)}|0}\to
\exp\left[-s_1s_2K_0\left(m\sqrt{-(x-y)^2+i\epsilon}\right)\right]
\label{dictB}
\end{equation}
\begin{equation}
\braket{0|e^{is_1\sqrt{2\pi} \,\cD_-(x)}\,e^{is_2\sqrt{2\pi} \,\cD_-(y)}|0}\to
(\xi m)^{s_1s_2}\,\left(-(x-y)^2+i\epsilon\right)^{s_1s_2/2}
\label{dictD}
\end{equation}
and the canonical dimensions are made up with factors of $\xi m/(2\pi)$.  So for
example, in (\ref{12}), one factor of  $\xi m/(2\pi)$ comes
from each of the two operators in the correlator, a $1/(\xi m)$ comes from
(\ref{dictD}).

Note that the results of this dictionary are identical to those of
\cite{Georgi:2020jik} but there they were derived in a much more
complicated way by first perturbatively evaluating correlators involving
operators with zero
dimension and then using cluster decomposition to isolate the
nonperturbative contributions.  The dictionary, (\ref{tempted}),
(\ref{dictB}), and (\ref{dictD}), does all
this automatically as long as we only apply it to the non-zero correlators
with zero chiral $T_3$.

Notice also that a parity transformation interchanges 
\begin{equation}
O_1\leftrightarrow O_1^*
\mbox{~~~and~~~}
O_2\leftrightarrow O_2^*
\label{parity}
\end{equation}
so the $\cB$ and $\cD_-$ fields in (\ref{tempted}) are pseudo-scalars.

\section{Conformal coalescence, parity, and $\pm$ mass \label{sec-ccmass}}

Notice that in (\ref{tempted}), the pair of operators $O_1$ and $O_2^*$
(and similarly the conjugate  pair $O_1^*$ and $O_2$) have the same dependence
on $\cD_-$.  We are also interested in the parity, so we define the operators
\begin{equation}
\begin{array}{l}
\displaystyle
O_{+-}=\frac{1}{2}\left(\frac{\xi m}{2\pi}\right)\Bigl(O_1-O_2^*+O_1^*-O_2\Bigr)
\\
\displaystyle
O_{--}=\frac{1}{2i}\left(\frac{\xi m}{2\pi}\right)\Bigl(O_1-O_2^*-O_1^*+O_2\Bigr)
\\
\displaystyle
O_{++}=\frac{1}{2}\left(\frac{\xi m}{2\pi}\right)\Bigl(O_1+O_2^*+O_1^*+O_2\Bigr)
\\
\displaystyle
O_{-+}=\frac{1}{2i}\left(\frac{\xi m}{2\pi}\right)\Bigl(O_1+O_2^*-O_1^*-O_2\Bigr)
\\
\end{array}
\label{pandl}
\end{equation} 
where the first subscript gives the parity and the second subscript
controls the low-energy behavior.
Every term in the expansion of the $O_{+-}$ and $O_{--}$ operators contains at
least one massive $\cB$.  Thus we expect these operators to disappear from
the low-energy theory and we expect the $O_{++}$ and $O_{-+}$ operators to
simplify.
At low energies we expect (always with $\theta=0$)
\begin{equation}
\begin{array}{l}
\displaystyle
O_{+-}
=i\left(\frac{\xi m}{2\pi}\right)\sin\Bigl(\sqrt{2\pi}\,\cB\Bigr)\,\left(e^{i\sqrt{2\pi}\,\cD_-}-e^{-i\sqrt{2\pi}\,\cD_-}\right)
\to0
\\
\displaystyle
O_{--}
=\left(\frac{\xi m}{2\pi}\right)\sin\Bigl(\sqrt{2\pi}\,\cB\Bigr)\,\left(e^{i\sqrt{2\pi}\,\cD_-}+e^{-i\sqrt{2\pi}\,\cD_-}\right)
\to0
\\
\displaystyle
O_{++}
=\left(\frac{\xi m}{2\pi}\right)\cos\Bigl(\sqrt{2\pi}\,\cB\Bigr)\,\left(e^{i\sqrt{2\pi}\,\cD_-}+e^{-i\sqrt{2\pi}\,\cD_-}\right)
\\
\displaystyle
\to
\left(\frac{\xi m}{2\pi}\right)\left(e^{i\sqrt{2\pi}\,\cD_-}+e^{-i\sqrt{2\pi}\,\cD_-}\right)
\
\\
\displaystyle
O_{-+}
=-i\left(\frac{\xi m}{2\pi}\right)\cos\Bigl(\sqrt{2\pi}\,\cB\Bigr)
\,\left(e^{i\sqrt{2\pi}\,\cD_-}-e^{-i\sqrt{2\pi}\,\cD_-}\right)
\\
\displaystyle
\to
-i\left(\frac{\xi m}{2\pi}\right)\left(e^{i\sqrt{2\pi}\,\cD_-}-e^{-i\sqrt{2\pi}\,\cD_-}\right)
\end{array}
\label{cc}
\end{equation} 
Note that it looks like we can combine the exponentials of $\pm
i\sqrt{2\pi}\,\cD_-$ into sines and cosines, but this would actually be
a mistake because these exponentials carry opposite
values of the chiral $T_3$ and must be treated separately in correlators.

The disappearance of the $O_{\pm -}$ fields in the low-energy limit was
discussed (in a more complicated way) in 
\cite{Georgi:2019tch} and \cite{Georgi:2020jik} and called ``conformal
coalescence''.  

The low-energy limits of the  non-zero 2-point functions of these fields
for zero fermion mass are exactly as expected from the low-energy forms in
(\ref{cc}).   
\begin{equation}
\begin{array}{c}
\displaystyle
\braket{0|T\,
O_{+-}(x)\,
O_{+-}(y)
|0}_0
=
\braket{0|T\,
O_{--}(x)\,
O_{--}(y)
|0}_0
\\
\displaystyle
=2\frac{(\xi
m)}{(2\pi)^2}
\sinh
\left[
K_0\left(m\sqrt{-(x-y)^2 + i\epsilon}\right)\right]
\,\left(-(x-y)^2+i\epsilon\right)^{-1/2}\to0
\end{array}
\label{--0}
\end{equation}
\begin{equation}
\begin{array}{c}
\displaystyle
\braket{0|T\,
O_{++}(x)\,
O_{++}(y)
|0}_0
=
\braket{0|T\,
O_{-+}(x)\,
O_{-+}(y)
|0}_0
\\
\displaystyle
=2\frac{(\xi
m)}{(2\pi)^2}
\cosh
\left[
K_0\left(m\sqrt{-(x-y)^2 + i\epsilon}\right)\right]
\,\left(-(x-y)^2+i\epsilon\right)^{-1/2}
\\
\displaystyle
\to2\frac{(\xi
m)}{(2\pi)^2}
\,\left(-(x-y)^2+i\epsilon\right)^{-1/2}
\end{array}
\label{++0}
\end{equation}
Thus we can calculate the low-energy correlators directly using
the low-energy forms.

We will use this to investigate the effect of a VERY SPECIAL fermion mass term.
We add to the Lagrangian (\ref{2f}) (for $n=2$) the fermion mass term
\begin{equation}
\delta\cL
=-\mu\Bigl(\bar\psi_1\psi_1-\bar\psi_2\psi_2\biggr)
=-2\mu\,O_{+-}
=2i\mu
\left(\frac{\xi
m}{2\pi}\right)\sin\Bigl(\sqrt{2\pi}\,\cB\Bigr)\,\left(e^{i\sqrt{2\pi}\,\cD_-}
-e^{-i\sqrt{2\pi}\,\cD_-}\right)
\label{o-mass}
\end{equation}
with equal and opposite masses for the fermions at $\theta=0$.
In \cite{Georgi:2020jik}, I briefly discussed the
consequences of a mass term like (\ref{o-mass}) proportional to $O_{+-}$.
Here I
will expand on this and 
calculate the matching of such a mass term onto the low-energy conformal
theory in a perturbation expansion in the fermion mass.  Normally, one
might expect a mass to produce a mass gap, eliminating the low-energy
conformal sector and breaking the conformal stuff into ordinary particles.
If this happens,
perturbation theory in the mass the parameter would be  
plagued by infrared divergences.  But in this case, because of the special
properties of the $O_{+-}$ operator, the matching occurs at the scale $m$
and the matching contribution involves only 
short distance physics.\footnote{See section 3 of \cite{Georgi:1993mps}.}
I will argue that this modifies the conformal symmetry without breaking it
while producing non-trivial interactions for the massive scalar and the
unparticle stuff..

The leading contribution at low energies is the second order term obtained
by integrating out the $\cB$, using
\begin{equation}
\braket{0|\cB(z_1)\,\cB(z_2)|0}\to -\frac{i}{m^2}\,\delta(z_1-z_2)
\end{equation}
which gives an effective interaction 
\begin{equation}
\left(\frac{\xi^2\mu^2}{\pi}\right)\left(e^{2i\sqrt{2\pi}\,\cD_-}
+e^{-2i\sqrt{2\pi}\,\cD_-}-2\right)
\label{mu2}
\end{equation}

The fermion mass term (\ref{o-mass}) breaks the chiral symmetry but
not parity so operators with different parity do not mix.  Thus we are interested
in the diagonal correlators in the low-energy effective theory below the $m$ scale,
\begin{equation}
\begin{array}{c}
\displaystyle
\braket{0|T\,
O_{\pm +}(x)\,
O_{\pm +}(y)
|0}_\mu
\\
\displaystyle
=
\pm\left(\frac{\xi m}{2\pi}\right)^2
\braket{0|T
\left(e^{i\sqrt{2\pi}\,\cD_-(x)}\pm e^{-i\sqrt{2\pi}\,\cD_-(x)}\right)
\left(e^{i\sqrt{2\pi}\,\cD_-(y)}\pm e^{-i\sqrt{2\pi}\,\cD_-(y)}\right)
|0}_\mu
\end{array}
\label{++mu}
\end{equation}

The first-order term in $\mu^2$ is
\begin{equation}
\begin{array}{c}
\displaystyle
\pm i\,\left(\frac{\xi m}{2\pi}\right)^2\left(\frac{\xi^2\mu^2}{\pi}\right)
\int\,d^2z
\\
\displaystyle
\bra{0}
T\,\left(e^{i\sqrt{2\pi}\,\cD_-(x)}\pm e^{-i\sqrt{2\pi}\,\cD_-(x)}\right)
\left(e^{i\sqrt{2\pi}\,\cD_-(y)}\pm e^{-i\sqrt{2\pi}\,\cD_-(y)}\right)
\\
\displaystyle
\left(e^{2i\sqrt{2\pi}\,\cD_-(z)}
+e^{-2i\sqrt{2\pi}\,\cD_-(z)}-2\right)
\ket{0}_0
\end{array}
\label{d++*mu2}
\end{equation}
We can now evaluate this by looking for the terms with chiral $T_3=0$.

The third term in the third set of parentheses is not interesting.
Because it doesn't
depend on $z$, it is just a vacuum energy contribution (which is the same
for both the $O_{++}$ and $O_{-+}$ correlators 
as it must be) and  so we can ignore it.
But the first and second terms give non-trivial contributions - both the
same so they add with the result
\begin{equation}
\pm i\mu^2\,\frac{\xi}{2m\pi^3}\,\int\,d^2z\,
\frac{\sqrt{-(x-y)^2+i\epsilon}}
{
\Bigl(-(x-z)^2+i\epsilon\Bigr)
\Bigl(-(y-z)^2+i\epsilon\Bigr)
}
\label{d++mu2-l2}
\end{equation}
Thus we need the integral
\begin{equation}
\int\,d^2z\,
\frac{1}
{
\Bigl(-(x-z)^2+i\epsilon\Bigr)
\Bigl(-(y-z)^2+i\epsilon\Bigr)
}
\label{2pt-int}
\end{equation}

Similar integrals in momentum space are very familiar but here the roles of UV
and IR divergences are reversed!  
There are short-distance singularities at
$z=x$ and $z=y$ as expected, because the non-zero mass term requires
regularization and is multiplicatively renormalized.\footnote{This is simply
related to the subtractive regularization of the 2-point function of the
fermion-bilinears}. But there is no large $z$ infrared divergence
so the expansion 
in powers of $\mu$ makes sense.  
We can combine denominators as usual to get
\begin{equation}
\int_0^1d\alpha\int\,d^{2}z\,
\frac{1}
{
\Bigl(-z^2+\alpha(1-\alpha)(-(x-y)^2+i\epsilon)\Bigr)^2
}
\label{d++mu2-l2r1}
\end{equation}
Wick rotation is now $z^0\to-iz^2$ and the integral becomes the Euclidean integral
\begin{equation}
-i
\int_0^1d\alpha\int\,d^{2}z\,
\frac{1}
{
\Bigl(z^2+\alpha(1-\alpha)(-(x-y)^2)\Bigr)^2
}
\label{d++mu2-l2r2}
\end{equation}

One way to deal with the short distance
singularities is to use dimensional regularization.  Because we are
computing a matching contribution onto the long-distance theory for
distances larger than $1/m$, it is appropriate to choose the
dimensional scale to be the matching scale of order $m$.  
Then our integral becomes (in dimension $2+\eta$)
\begin{equation}
-i\,m^{\eta}\int_0^1d\alpha\int\,d^{2+\eta}z\,
\frac{\sqrt{-(x-y)^2}}
{
\Bigl(z^2+\alpha(1-\alpha)(-(x-y)^2)\Bigr)^2
}
\label{d++mu2-l2r2e}
\end{equation}
\begin{equation}
=-i\,
m^{\eta}\frac{2\pi^{1+\eta/2}}{\Gamma(1+\eta/2)}
\,\int_0^1d\alpha\int_0^\infty\,dz\,z^{1-\eta}\,
\frac{\sqrt{-(x-y)^2}}
{
\Bigl(z^2+\alpha(1-\alpha)(-(x-y)^2)\Bigr)^2
}
\label{d++mu2-l2r3}
\end{equation}
\begin{equation}
=-i\,
m^{\eta}\frac{2\pi^{1+\eta/2}}{\Gamma(1+\eta/2)}
\,\frac{\pi\eta}{4\sin(\pi\eta/2)}\,\int_0^1d\alpha\,
\frac{\sqrt{-(x-y)^2}}
{
\Bigl(\alpha(1-\alpha)(-(x-y)^2\Bigr)^{1-\eta/2}
}
\label{d++mu2-l2r4}
\end{equation}
\begin{equation}
=-i\,
m^{\eta}
\frac{2\pi^{1+\eta/2}}{\Gamma(1+\eta/2)}
\,\frac{\pi\eta}{4\sin(\pi\eta/2)}
\,\frac{\Gamma(\eta/2)}{\Gamma(\eta)}\,
\frac{\sqrt{-(x-y)^2+i\eta}}
{
\Bigl((-(x-y)^2+i\eta\Bigr)^{1-\eta/2}
}
\label{d++mu2-l2r5}
\end{equation}
Expanding the result in powers of $\eta$ and putting back the original
$i \epsilon$ for Minkowski space gives
\begin{equation}
\int\,d^2z\,
\frac{1}
{
\Bigl(-(x-z)^2+i\epsilon\Bigr)
\Bigl(-(y-z)^2+i\epsilon\Bigr)
}
=-4i\pi\,\frac{\log\Bigl(m\sqrt{-(x-y)^2+i\epsilon}\Bigr)}
{\Bigl(-(x-y)^2+i\epsilon\Bigr)}
\label{2pt-int-r}
\end{equation}
so (\ref{d++mu2-l2}) becomes
\begin{equation}
\pm\mu^2\,\frac{2\xi}{\pi^2 m}\frac{\log\Bigl(m\sqrt{-(x-y)^2+i\epsilon}\Bigr)}
{\sqrt{-(x-y)^2+i\epsilon}}
\label{d++mu2-l2r6}
\end{equation}

Dimensional regularization works simply enough in this case, but it will be
useful to understand the integral in different ways.  The
long-distance behavior should be independent of the details of our short
distance regularization. In particular, we can cut off the short distance
behavior in (\ref{d++mu2-l2r1}) at $1/m$ by adding a $1/m^2$ term to get
\begin{equation}
-i\,\int_0^1d\alpha\int\,d^{2}z\,
\frac{\sqrt{-(x-y)^2}}
{
\Bigl(z^2+\alpha(1-\alpha)(-(x-y)^2)+1/m^2\Bigr)^2
}
\label{d++mu2-l2r2m}
\end{equation}
\begin{equation}
=-i\pi\,\int_0^1d\alpha\,
\frac{1}
{
\Bigl(\alpha(1-\alpha)(-(x-y)^2)+1/m^2\Bigr)
}
\label{d++mu2-l2r2m2}
\end{equation}
In the long-distance limit, $-(x-y)^2\gg1/m^2$, this again gives (\ref{2pt-int-r}).

The key things in (\ref{d++mu2-l2r6}) are the appearance of the
logarithm of $m$ and the absence of any logarithm of $\mu$.  This
again shows
that the matching is happening at the scale $m$ and there are no IR divergences.   
The log of $m$ does not indicate that conformal invariance is broken.
Rather, it is exactly what we would expect
if the conformal symmetry of the 2-point functions 
at long distances is not broken to this order in $\mu^2$ but the
parity eigenstate operators, $O_{\pm +}$ have scaling
dimensions $d_{\pm}$ that change in opposite directions when the mass term
is turned on. 
Adding (\ref{d++mu2-l2r6}) to the zeroth-order contribution gives
\begin{equation}
\frac{\xi
m}{2\pi^2}
\,\frac{1}
{\sqrt{-(x-y)^2+i\epsilon}}
\pm\mu^2\,\frac{2\xi}{\pi^2 m}\frac{\log\Bigl(m\sqrt{-(x-y)^2+i\epsilon}\Bigr)}
{\sqrt{-(x-y)^2+i\epsilon}}
\end{equation}
which is the expansion 
to leading nontrivial order in $\mu$ of
\begin{equation}
\displaystyle
\braket{0|T\,
O_{\pm +}(x)\,
O_{\pm +}(y)
|0}_\mu
=\xi\,\frac{m^{2-2d_{\pm}}}{2\pi^2}
\,\frac{1}{\left(-(x-y)^2+i\epsilon\right)^{d_\pm}}
\label{++beta}
\end{equation}
where
\begin{equation}
d_\pm=\frac{1\mp 4\mu^2/m^2}{2}
\label{d+-}
\end{equation}

\section{4-point functions\label{sec-4pt}}

The non-zero 4-point functions in the massless theory at long distances 
can be calculated
simply from (\ref{dictD}) and (\ref{cc}). They are\footnote{Note the
particular pattern of $O_{++}$s and $O_{-+}$s in (\ref{+-+-}).  This is
just a convenience to make the result easy to write down compactly.}
\begin{equation}
\begin{array}{c}
\displaystyle
\braket{0|T\,
O_{\pm +}(x_1)\,
O_{\pm +}(x_2)\,
O_{\pm +}(x_3)\,
O_{\pm +}(x_4)
|0}_0
=2\left(\frac{\xi m}{2\pi}\right)^4\,\frac{1}{(\xi m)^2}\,
\sum_{3\;\not =\;\rm{perms}\atop\{jklm\}=\{1234\}}
\\
\displaystyle
\rule{0pt}{7.5ex}
\sqrt{\frac{\Bigl(-(x_j-x_k)^2+i\epsilon\Bigr)\Bigl(-(x_l-x_m)^2+i\epsilon\Bigr)}
{\Bigl(-(x_j-x_l)^2+i\epsilon\Bigr)\Bigl(-(x_j-x_m)^2+i\epsilon\Bigr)
\Bigl(-(x_k-x_l)^2+i\epsilon\Bigr)\Bigl(-(x_k-x_m)^2+i\epsilon\Bigr)}}
\end{array}
\label{++++}
\end{equation}
\begin{equation}
\begin{array}{c}
\displaystyle
\braket{0|T\,
O_{++}(x_1)\,
O_{-+}(x_2)\,
O_{++}(x_3)\,
O_{-+}(x_4)
|0}_0
\\
\displaystyle
=-2\left(\frac{\xi m}{2\pi}\right)^4\,\frac{1}{(\xi m)^2}\,
\sum_{3\;\not =\;\rm{perms}\atop\{jklm\}=\{1234\}}(-1)^{j+k}
\\
\displaystyle
\rule{0pt}{7.5ex}
\sqrt{\frac{\Bigl(-(x_j-x_k)^2+i\epsilon\Bigr)\Bigl(-(x_l-x_m)^2+i\epsilon\Bigr)}
{\Bigl(-(x_j-x_l)^2+i\epsilon\Bigr)\Bigl(-(x_j-x_m)^2+i\epsilon\Bigr)
\Bigl(-(x_k-x_l)^2+i\epsilon\Bigr)\Bigl(-(x_k-x_m)^2+i\epsilon\Bigr)}}
\end{array}
\label{+-+-}
\end{equation}

We begin by discussing (\ref{++++}).
The order $\mu^2$ correction to (\ref{++++}) is
\begin{equation}
\begin{array}{c}
\displaystyle
i\,\left(\frac{\xi m}{2\pi}\right)^4\left(\frac{\xi^2\mu^2}{\pi}\right)
\int\,d^2z
\\
\displaystyle
\bra{0}
T\,\left(e^{i\sqrt{2\pi}\,\cD_-(x_1)}\pm e^{-i\sqrt{2\pi}\,\cD_-(x_1)}\right)
\left(e^{i\sqrt{2\pi}\,\cD_-(x_2)}\pm e^{-i\sqrt{2\pi}\,\cD_-(x_2)}\right)
\\
\displaystyle
\left(e^{i\sqrt{2\pi}\,\cD_-(x_3)}\pm e^{-i\sqrt{2\pi}\,\cD_-(x_3)}\right)
\left(e^{i\sqrt{2\pi}\,\cD_-(x_4)}\pm e^{-i\sqrt{2\pi}\,\cD_-(x_4)}\right)
\\
\displaystyle
\left(e^{2i\sqrt{2\pi}\,\cD_-(z)}
+e^{-2i\sqrt{2\pi}\,\cD_-(z)}-2\right)
\ket{0}_0
\end{array}
\label{d++++mu2}
\end{equation}

As for the 2-point function, the third term in the last line gives an
irrelevant vacuum energy contribution while the first and second terms give
effects of the following form:
\begin{equation}
\pm i\mu^2\,\frac{\xi^6m^4}{8\pi^5}\,\int\,d^2z\,
\braket{0|T\,
e^{\pm i\sqrt{2\pi}\,\cD(x_j)}\,
e^{\pm i\sqrt{2\pi}\,\cD(x_k)}\,
e^{\pm i\sqrt{2\pi}\,\cD(x_l)}\,
e^{\mp i\sqrt{2\pi}\,\cD(x_m)}
e^{\mp 2i\sqrt{2\pi}\,\cD(z)}
|0}_0
\label{d++++*mu2-l}
\end{equation}
(the $j$, $k$ and $l$ indices have the same sign in the exponent) which gives
\begin{equation}
\pm i\mu^2\,\frac{\xi^2}{8\pi^5}\,\sum_m\int\,d^2z\,
\frac{
\Bigl(-(z-x_m)^2+i\epsilon\Bigr)\,
\prod_{j<k\atop j,k\neq m}
\sqrt{-(x_j-x_k)^2+i\epsilon}
}
{
\prod_{n=\atop j,k,l}
\Bigl(-(z-x_n)^2+i\epsilon\Bigr)\,
\sqrt{-(x_n-x_m)^2+i\epsilon}
}
\label{d++++*mu2-l2}
\end{equation}

So we need the integral\footnote{The $i\epsilon$ is not necessary in the numerator.}
\begin{equation}
\int\,d^2z\,
\frac{
\Bigl(-(z-x_m)^2\Bigr)\,
}
{
\prod_{n=\atop j,k,l}
\Bigl(-(z-x_n)^2+i\epsilon\Bigr)\,
}
\label{d++++*int1}
\end{equation}
\begin{equation}
=2\int\,[d\alpha]\,d^2z\,
\delta\left(1-\sum_{n=\atop j,k,l}\alpha_n\right)\,
\frac{
\Bigl(-(z-x_m)^2\Bigr)\,
}
{
\left(\sum_{n=\atop j,k,l}\alpha_n
\Bigl(-(z-x_n)^2+i\epsilon\Bigr)\right)^3\,
}
\label{d++++*int2}
\end{equation}
\begin{equation}
=2\int\,[d\alpha]\,d^2z\,
\delta\left(1-\sum_{n=\atop j,k,l}\alpha_n\right)\,
\frac{
\Bigl(-(z-x_m)^2\Bigr)\,
}
{
\left(-z^2+\sum_{n=\atop j,k,l}\Bigl(2\alpha_n (zx_n)-\alpha_nx_n^2\Bigr)
+i\epsilon\right)^3\,
}
\label{d++++*int3}
\end{equation}
\begin{equation}
=2\int\,[d\alpha]\,d^2\tilde z\,
\delta\left(1-\sum_{n=\atop j,k,l}\alpha_n\right)\,
\frac{
\left(-\left(\tilde z-\Bigl(\sum_{n=\atop j,k,l}\alpha_nx_n\Bigr)-x_m\right)^2\right)\,
}
{
\left(-{\tilde z}^2-\Bigl(\sum_{n=\atop j,k,l}\alpha_nx_n^2\Bigr)
+\Bigl(\sum_{n=\atop j,k,l}\alpha_nx_n\Bigr)^2+i\epsilon\right)^3\,
}
\label{d++++*int3'}
\end{equation}
\begin{equation}
=2\int\,[d\alpha]\,d^2z\,
\delta\left(1-\sum_{n=\atop j,k,l}\alpha_n\right)\,
\frac{
\left(-z^2-\left(x_m+\left(\sum_{n=\atop j,k,l}\alpha_nx_n\right)\right)^2\right)\,
}
{
\left(-z^2-\left(\sum_{n=\atop j,k,l}\Bigl(\alpha_n x_n^2\Bigr)\right)
+\left(\sum_{n=\atop j,k,l}\alpha_nx_n\right)^2+i\epsilon\right)^3\,
}
\label{d++++*int4}
\end{equation}

Now we can Wick rotate and do the $z$ integration
\begin{equation}
\begin{array}{c}
\displaystyle
=-2i
\int\,[d\alpha]\,d^{2}z\,
\delta\left(1-\sum_{n=\atop j,k,l}\alpha_n\right)\,
\frac{
\left(z^2+a\right)\,
}
{
\left(z^2+b\right)^3\,
}
\\
\displaystyle
=-i\pi\,\int\,[d\alpha]\,
\delta\left(1-\sum_{n=\atop j,k,l}\alpha_n\right)\,
\frac{b+a}{b^2}
\end{array}
\label{d++++*int5}
\end{equation}
where
\begin{equation}
a=-\left(x_m-\left(\sum_{n=\atop j,k,l}\alpha_nx_n\right)\right)^2
\end{equation}
\begin{equation}
b=\left(\sum_{j=1}^3\Bigl(-\alpha_j x_j^2\Bigr)\right)
+\left(\sum_{n=\atop j,k,l}\alpha_nx_n\right)^2
\end{equation}
and because $\sum_{n=j,k,l}\alpha_n=1$ in the integral, we can write
\begin{equation}
b=-\sum_{j<k}\alpha_j\alpha_k(x_j-x_k)^2
\end{equation}
\begin{equation}
b+a=-\sum_{n=\atop j,k,l}\alpha_n(x_m-x_n)^2
\end{equation}
If we cut off the integral at short distance as in
(\ref{d++mu2-l2r2m}), (\ref{d++++*int5}) becomes
\begin{equation}
=-i\pi\,\int\,[d\alpha]\,
\delta\left(1-\sum_{n=\atop j,k,l}\alpha_n\right)\,
\frac{\left(-\sum_{n=\atop j,k,l}\alpha_n(x_m-x_n)^2\right)}
{\left(-\sum_{j<k}\alpha_j\alpha_k(x_j-x_k)^2+1/m^2\right)}
\end{equation}
In the long-distance limit, $-(x_j-x_k)^2\gg1/m^2$, this gives (suppressing
the $i\epsilon$s in the result)
\begin{equation}
\begin{array}{c}
\displaystyle
\int\,d^2z\,
\frac{
\Bigl(-(z-x_m)^2\Bigr)\,
}
{
\prod_{n=\atop j,k,l}
\Bigl(-(z-x_n)^2+i\epsilon\Bigr)
}
\\
\displaystyle
\to -2i\pi\sum_{j\neq m}\frac{-(x_m-x_j)^2}{(x_k-x_j)^2(x_l-x_j)^2}
\log\left(m\sqrt{-(x_k-x_j)^2}\sqrt{-(x_l-x_j)^2}/\sqrt{-(x_k-x_l)^2}\right)
\end{array}
\label{4pt-int}
\end{equation}

Putting all this together gives
\begin{equation}
\begin{array}{c}
\displaystyle
\pm \mu^2\,\frac{\xi^2}{4\pi^4}\,\sum_m
\frac{
\prod_{j<k\atop j,k\neq m}
\sqrt{-(x_j-x_k)^2}
}
{
\prod_{n=\atop j,k,l}
\sqrt{-(x_n-x_m)^2}
}
\\
\displaystyle
\sum_{j\neq m}\frac{-(x_m-x_j)^2}{(x_k-x_j)^2(x_l-x_j)^2}
\log\left(m\sqrt{-(x_k-x_j)^2}\sqrt{-(x_l-x_j)^2}/\sqrt{-(x_k-x_l)^2}\right)
\end{array}
\label{d++++*mu2-l2f}
\end{equation}
\begin{equation}
\begin{array}{c}
\displaystyle
=\pm \mu^2\,\frac{\xi^2}{2\pi^4}\,
\sum_{{\rm pairs\;}\{j,k\}=\atop\{1,2\},\{1,3\},\{1,4\}}
\frac{
\sqrt{-(x_j-x_k)^2}
\sqrt{-(x_l-x_m)^2}
}
{
\sqrt{-(x_j-x_l)^2}
\sqrt{-(x_j-x_m)^2}
\sqrt{-(x_k-x_l)^2}
\sqrt{-(x_k-x_m)^2}
}
\\
\displaystyle
\log\left(
\frac{m^4\sqrt{-(x_j-x_l)^2}
\sqrt{-(x_j-x_m)^2}
\sqrt{-(x_k-x_l)^2}
\sqrt{-(x_k-x_m)^2}
}{
\sqrt{-(x_j-x_k)^2}
\sqrt{-(x_l-x_m)^2}
}
\right)
\end{array}
\label{d++++*mu2-l2f2}
\end{equation}
Adding the zeroth order term gives
\begin{equation}
\begin{array}{c}
\displaystyle
2\left(\frac{\xi m}{2\pi}\right)^4\left(\frac{1}{\xi m}\right)^2
\sum_{{\rm pairs\;}\{j,k\}=\atop\{1,2\},\{1,3\},\{1,4\}}
\frac{
\sqrt{-(x_j-x_k)^2}
\sqrt{-(x_l-x_m)^2}
}
{
\sqrt{-(x_j-x_l)^2}
\sqrt{-(x_j-x_m)^2}
\sqrt{-(x_k-x_l)^2}
\sqrt{-(x_k-x_m)^2}
}
\\
\displaystyle
\left(1\pm\frac{4\mu^2}{m^2}
\log\left(
\frac{m^4\sqrt{-(x_j-x_l)^2}
\sqrt{-(x_j-x_m)^2}
\sqrt{-(x_k-x_l)^2}
\sqrt{-(x_k-x_m)^2}
}{
\sqrt{-(x_j-x_k)^2}
\sqrt{-(x_l-x_m)^2}
}
\right)
\right)
\end{array}
\label{d++++*mu2-l2f3}
\end{equation}
This is the expansion to order $\mu^2$ of
\begin{equation}
\frac{\xi^2m^{4-4d_{\pm}}}{8\pi^4}
\sum_{{\rm pairs\;}\{j,k\}=\atop\{1,2\},\{1,3\},\{1,4\}}
\frac{
\left(-(x_j-x_k)^2\right)^{d_\pm}
\left(-(x_l-x_m)^2\right)^{d_\pm}
}
{
\left(-(x_j-x_l)^2\right)^{d_\pm}
\left(-(x_j-x_m)^2\right)^{d_\pm}
\left(-(x_k-x_l)^2\right)^{d_\pm}
\left(-(x_k-x_m)^2\right)^{d_\pm}
}
\end{equation}
This is consistent with the result (\ref{++beta}) for the 2-point function
and is the simplest possible result consistent
with conformal invariance for the $O_{\pm +}$ fields with scaling
dimensions $d_{\pm}$. 

In the mixed correlator, (\ref{+-+-}), the pieces of the calculation are
very similar but they get put together very differently. 
The first order correction to (\ref{+-+-}) is
\begin{equation}
\begin{array}{c}
\displaystyle
-i\mu^2\,\frac{\xi^6m^4}{8\pi^5}\,\int\,d^2z\,
\Bigl(\braket{0|T\,
e^{- i\sqrt{2\pi}\,\cD(x_1)}\,
e^{ i\sqrt{2\pi}\,\cD(x_2)}\,
e^{ i\sqrt{2\pi}\,\cD(x_3)}\,
e^{i\sqrt{2\pi}\,\cD(x_4)}
e^{-2i\sqrt{2\pi}\,\cD(z)}
|0}_0
\\
\displaystyle
-\braket{0|T\,
e^{ i\sqrt{2\pi}\,\cD(x_1)}\,
e^{- i\sqrt{2\pi}\,\cD(x_2)}\,
e^{ i\sqrt{2\pi}\,\cD(x_3)}\,
e^{i\sqrt{2\pi}\,\cD(x_4)}
e^{-2i\sqrt{2\pi}\,\cD(z)}
|0}_0
\\
\displaystyle
+\braket{0|T\,
e^{ i\sqrt{2\pi}\,\cD(x_1)}\,
e^{ i\sqrt{2\pi}\,\cD(x_2)}\,
e^{- i\sqrt{2\pi}\,\cD(x_3)}\,
e^{i\sqrt{2\pi}\,\cD(x_4)}
e^{-2i\sqrt{2\pi}\,\cD(z)}
|0}_0
\\
\displaystyle
-\braket{0|T\,
e^{ i\sqrt{2\pi}\,\cD(x_1)}\,
e^{ i\sqrt{2\pi}\,\cD(x_2)}\,
e^{ i\sqrt{2\pi}\,\cD(x_3)}\,
e^{-i\sqrt{2\pi}\,\cD(x_4)}
e^{-2i\sqrt{2\pi}\,\cD(z)}
|0}_0\Bigr)
\end{array}
\label{d+-+-mu2-l}
\end{equation}
\begin{equation}
=i\mu^2\,\frac{\xi^2}{8\pi^5}\,\sum_m(-1)^m\int\,d^2z\,
\frac{
\Bigl(-(z-x_m)^2\Bigr)\,
\prod_{j<k\atop j,k\neq m}
\sqrt{-(x_j-x_k)^2+i\epsilon}
}
{
\prod_{n=\atop j,k,l}
\Bigl(-(z-x_n)^2+i\epsilon\Bigr)\,
\sqrt{-(x_n-x_m)^2+i\epsilon}
}
\label{d+-+-*mu2-l2}
\end{equation}
Except for a factor of
$(-1)^m$ in the sum, this is proportional to (\ref{d++++*mu2-l2}). 
So following the same steps and putting everything together now gives	
\begin{equation}
\begin{array}{c}
\displaystyle
\mu^2\,\frac{\xi^2}{4\pi^4}\,\sum_m(-1)^m
\frac{
\prod_{j<k\atop j,k\neq m}
\sqrt{-(x_j-x_k)^2}
}
{
\prod_{n=\atop j,k,l}
\sqrt{-(x_n-x_m)^2}
}
\sum_{j\neq m}\frac{-(x_m-x_j)^2}{(x_k-x_j)^2(x_l-x_j)^2}
\\
\displaystyle
\log\left(m\sqrt{-(x_k-x_j)^2}\sqrt{-(x_l-x_j)^2}/\sqrt{-(x_k-x_l)^2}\right)
\end{array}
\label{d++++*mu2-l2f4}
\end{equation}
\begin{equation}
\begin{array}{c}
\displaystyle
=\mu^2\,\frac{\xi^2}{4\pi^4}\,
\left(\prod_{\alpha\neq\beta}\frac{1}{\sqrt{-(x_\alpha-x_\beta)^2}}\right)
\left(\rule{0pt}{4ex}\right.\Bigl(-(x_1-x_3)^2\Bigr)\Bigl(-(x_2-x_4)^2\Bigr)
\log\left(\frac{-(x_2-x_4)^2}{-(x_1-x_3)^2}\right)
\\
\displaystyle
-\Bigl(-(x_1-x_2)^2\Bigr)\Bigl(-(x_3-x_4)^2\Bigr)
\log\left(\frac{-(x_2-x_4)^2}{-(x_1-x_3)^2}\right)
\\
\displaystyle
-\Bigl(-(x_1-x_4)^2\Bigr)\Bigl(-(x_2-x_3)^2\Bigr)
\log\left(\frac{-(x_2-x_4)^2}{-(x_1-x_3)^2}\right)
\left.\rule{0pt}{4ex}\right)
\end{array}
\end{equation}
\begin{equation}
=\frac{2\mu^2}{m^2}
\log\left(\frac{-(x_2-x_4)^2}{-(x_1-x_3)^2}\right)
\braket{0|T\,
O_{++}(x_1)\,
O_{-+}(x_2)\,
O_{++}(x_3)\,
O_{-+}(x_4)
|0}_0
\end{equation}
Adding the zeroth order term gives
\begin{equation}
\left(1+\frac{2\mu^2}{m^2}
\log\left(\frac{-(x_2-x_4)^2}{-(x_1-x_3)^2}\right)\right)
\braket{0|T\,
O_{++}(x_1)\,
O_{-+}(x_2)\,
O_{++}(x_3)\,
O_{-+}(x_4)
|0}_0
\end{equation}
which is the expansion of
\begin{equation}
\frac{\left(-(x_2-x_4)^2\right)^{2\mu^2/m^2}}
{\left(-(x_1-x_3)^2\right)^{2\mu^2/m^2}}
\braket{0|T\,
O_{++}(x_1)\,
O_{-+}(x_2)\,
O_{++}(x_3)\,
O_{-+}(x_4)
|0}_0
\end{equation}
So again conformal invariance is satisfied.

\section{Higher $2n$-point functions\label{sec-npt}}

The calculation of higher $2n$-point correlators is straightforward but quickly
gets complicated.  The leading correction to the $2n$-point function
involves integrals of the form
\begin{equation}
\int\,d^2z\,
\frac{
\prod_{j=n+2}^{2n}\Bigl(-(z-x_j)^2\Bigr)\,
}
{
\prod_{j=1}^{n+1}
\Bigl(-(z-x_j)^2+i\epsilon\Bigr)\,
}
\label{2n-int1}
\end{equation}
It would be a difficult task to calculate this using the direct methods of
section~\ref{sec-4pt}, but we can use (\ref{2pt-int-r}) and (\ref{4pt-int})
and linear algebra to write down the result without further integration.

Here is how this works for the the $6$-point function with the integral
\begin{equation}
\int\,d^2z\,
\frac{
\Bigl(-(z-x_5)^2\Bigr)\,\Bigl(-(z-x_6)^2\Bigr)
}
{
\prod_{j=1}^{n+1}
\Bigl(-(z-x_1)^2+i\epsilon\Bigr)\,
\Bigl(-(z-x_2)^2+i\epsilon\Bigr)\,
\Bigl(-(z-x_3)^2+i\epsilon\Bigr)\,
\Bigl(-(z-x_4)^2+i\epsilon\Bigr)
}
\label{6-int1}
\end{equation}
The point is that
the product 
\begin{equation}
\Bigl(-(z-x_5)^2\Bigr)\,\Bigl(-(z-x_6)^2\Bigr)
\label{56}
\end{equation}
can be written as a linear
combination of $\bigl(-(z-x_j)^2\bigr)\,\bigl(-(z-x_k)^2\bigr)$,
$\bigl(-(z-x_5)^2\bigr)\,\bigl(-(z-x_k)^2\bigr)$, and
$\bigl(-(z-x_j)^2\bigr)\,\bigl(-(z-x_6)^2\bigr)$  where $j$ and $k$ are less
than $5$.  In fact, this can be done in many different ways because the $z$
dependence can be written in terms of the nine combinations
\begin{equation}
\Bigl\{z_+^2,\,z_+,\,1\Bigr\}\times\Bigl\{z_-^2,\,z_-,\,1\Bigr\}
\end{equation}
where
\begin{equation}
z_\pm=z^0\pm z^1
\end{equation}
So we can write
\begin{equation}
\Bigl(-(z-x_5)^2\Bigr)\,\Bigl(-(z-x_6)^2\Bigr)
=\sum_{\{j,k\}\neq\{5,6\}}^{9\;\rm terms}\beta(j,k)
\Bigl(-(z-x_j)^2\Bigr)\,\Bigl(-(z-x_k)^2\Bigr)
\label{betasum}
\end{equation}
Then setting (\ref{56}) equal to (\ref{betasum}) gives 9 linear equations
for the 9 $\beta$s depending on the $x_{j\pm}$.
Then in each of the terms in the sum in (\ref{betasum}), one or two of the
factors cancel with factors in the denominator and the integral reduces to
(\ref{2pt-int-r}) or (\ref{4pt-int}).

A similar procedure can be used to calculate the leading corrections to the
$2n$-point function in terms of $n^2$ $\beta$s.

While this is simple to describe, I have not found a choice of $\beta$s
that leads to any simple or intuitive result.

\section{More questions\label{sec-more}}

The analysis above gives some nontrivial checks of the conjecture that the
2-flavor Schwinger model has an unbroken conformal sector even when small
equal and opposite fermion masses are turned on and describes the
calculation of the leading corrections to all flavor-diagonal correlators in a systematic
expansion in the fermion mass parameters..  But the simple
calculational scheme used here leaves
some questions unanswered.  
Can the matrix analysis of 6-point and higher correlators in
sections~\ref{sec-npt} be simplified.
What happens
in higher orders in $\mu^2$?  
Do the scaling
dimensions of the fermion-bilinears, (\ref{d+-}), 
give a clue to the nature of the phase transition that must occur between
$\mu^2\ll m^2$ and $\mu^2\gg m^2$?~\cite{Coleman:1976uz}  Are there
observable effects at high energies of the non-trivial dimensions in the
conformal sector? Can the analysis be extended to include other fermion
bilinears and the non-abelian chiral
symmetry.~\cite{Witten:1983ar}  And most 
importantly, does this solvable model give any
clue to how an \textbf{unbroken} conformal sector might show up in the particle
physics of our 3+1 dimensional world?\footnote{A referee suggested
\cite{Wong:2020hjc} for discussion of how a 1+1 dimensional model might be
relevant in 3+1 dimensions.} I believe that it is worth studying 
this model further. 

\section*{Acknowledgements\label{sec-ack}}

I am grateful for discussions with Tom Banks, Jacob Barandes, David Kaplan,
and for very important questions from
Hofie Hannesdottir and Rashmish Mishra.

This project has received support from the European Union's
Horizon 2020 research and innovation programme under the Marie 
Skodowska-Curie grant agreement No 860881-HIDDeN. 

\bibliography{up4}

\end{document}